\documentclass[conference]{IEEEtran}

\usepackage{algorithmic}
\usepackage{graphicx}
\usepackage{xcolor}
\usepackage{url}
\usepackage{multirow}
\usepackage{diagbox}
\usepackage{enumitem} 
\def\BibTeX{{\rm B\kern-.05em{\sc i\kern-.025em b}\kern-.08em
    T\kern-.1667em\lower.7ex\hbox{E}\kern-.125emX}}
\begin{document}

\title{Using a Nearest-Neighbour, BERT-Based Approach for Scalable Clone Detection}
\author{\IEEEauthorblockN{Muslim Chochlov}
\IEEEauthorblockA{\textit{Dept. of Computer Science and} \\
\textit{Information Systems} \\
\textit{University of Limerick}\\
Limerick, Ireland \\
muslim.chochlov@ul.ie}
\and
\IEEEauthorblockN{Gul Aftab Ahmed}
\IEEEauthorblockA{\textit{Dept. of Computer Science}\\
\textit{Trinity College Dublin} \\
Dublin, Ireland \\
ahmedga@tcd.ie}
\and
\IEEEauthorblockN{James Vincent Patten}
\IEEEauthorblockA{\textit{Dept. of Computer Science and} \\
\textit{Information Systems} \\
\textit{University of Limerick}\\
Limerick, Ireland \\
james.patten@ul.ie}
\and
\IEEEauthorblockN{Guoxian Lu}
\IEEEauthorblockA{\textit{WN Digital IPD and} \\
\textit{Trustworthiness Enabling} \\
\textit{Huawei Technologies Co., Ltd.}\\
Shanghai, China \\
luguoxian@huawei.com}
\and
\IEEEauthorblockN{Wei Hou}
\IEEEauthorblockA{\textit{Huawei Vulnerability} \\
\textit{Management Center} \\
\textit{Huawei Technologies Co., Ltd.}\\
Shenzhen, Guangdong, China \\
wei.hou@huawei.com}
\and
\IEEEauthorblockN{David Gregg}
\IEEEauthorblockA{\textit{Dept. of Computer Science}\\
\textit{Trinity College Dublin} \\
Dublin, Ireland \\
david.gregg@tcd.ie}
\and
\IEEEauthorblockN{Jim Buckley}
\IEEEauthorblockA{\textit{Dept. of Computer Science and} \\
\textit{Information Systems} \\
\textit{University of Limerick}\\
Limerick, Ireland \\
jim.buckley@ul.ie}
}
\hyphenation{SourcererCC}
\maketitle

\begin{abstract}
Code clones can detrimentally impact software maintenance and manually detecting them in very large codebases is impractical. Additionally, automated approaches find detection of Type 3 and Type 4 (inexact) clones very challenging. While the most recent artificial deep neural networks (for example BERT-based artificial neural networks) seem to be highly effective in detecting such clones, their pairwise comparison of every code pair in the target system(s) is inefficient and scales poorly on large codebases.

We therefore introduce SSCD, a BERT-based clone detection approach that targets high recall of Type 3 and Type 4 clones at scale (in line with our industrial partner’s requirements). It does so by computing a representative embedding for each code fragment and finding similar fragments using a nearest neighbour search. SSCD thus avoids the pairwise-comparison bottleneck of other Neural Network approaches while also using parallel, GPU-accelerated search to tackle scalability.

This paper details the approach and an empirical assessment towards configuring and evaluating that approach in industrial setting. The configuration analysis suggests that shorter input lengths and text-only based neural network models demonstrate better efficiency in SSCD, while only slightly decreasing effectiveness. The evaluation results suggest that SSCD is more effective than state-of-the-art approaches like SAGA and SourcererCC. It is also highly efficient: in its optimal setting, SSCD effectively locates clones in the entire 320 million LOC BigCloneBench (a standard clone detection benchmark) in just under three hours.
\end{abstract}

\begin{IEEEkeywords}
Clone detection, semantic clones, scalable, deep neural networks
\end{IEEEkeywords}

\section{Introduction}
Clone detection is the process of locating textually exact or similar pieces of code\cite{Rattan2013}. Clones can be classified as:

\begin{itemize}[leftmargin=*]
    \item Type 1 (T1): Identical except for whitespaces and comments.
    \item Type 2 (T2): Where variables and identifiers are renamed.
    \item Type 3 (T3): Additionally there are code re-orderings, insertions and deletions.
    \item Type 4 (T4): Functionally similar clones that may be textually very different \cite{Rattan2013}.
\end{itemize}

Locating clones manually is inefficient, and many automatic clone detection techniques (CDTs) have been proposed. Classical CDTs use data structures such as suffix trees to efficiently find T1 and T2 clones. But automatically finding T3/T4 clones is much more difficult, because the matching is no longer exact. Svajlenko et al.'s classification of T3 and T4 clones illustrates the approximate nature of the matching \cite{Svajlenko2014a}:
\begin{itemize}
    \item Very Strong T3 (VST3: 90\%-100\% textually similar)
    \item Strong T3 (ST3: 70\%-90\% textually similar)
    \item Medium T3 (MT3: 50\%-70\% textually similar)
    \item Weak T3/T4 (WT3/T4: \textless50\% textually similar)
\end{itemize}

Researchers have proposed many solutions to this approximate matching problem, such as representing code fragments as bags of tokens. SourcererCC adopts this approach \cite{Sajnani2016} but, even with filtering heuristics, T3/T4 clone detection is a difficult and computationally intensive task for it: it required over 4.5 days to complete clone detection on a 250+ MLOC version of the BigCloneBench benchmark (BCB)\cite{Svajlenko2014a}.

Recently, several artificial neural-network (NN) based approaches have been proposed to improve the accuracy of T3/T4 clone detection. These methods achieve higher recall (finding true clones), and better precision (rejecting false clones) than the best classical methods \cite{White2016, Wei2017, Saini2018, Buch2019, Zhang2019, Guo2021}. In these approaches, novel NNs such as ASTNN \cite{Zhang2019}, CodeBERT \cite{Feng2020}, and GraphCodeBERT \cite{Guo2021} are trained to recognize whether a given pair of code pieces are clones or not. However, this pairwise comparison approach has performance issues when faced with large codebases: If there are $n$ code fragments in the codebase, then $n(n-1)/2$ (that is $O(n^2)$) comparisons are required. Testing all these possible combinations is infeasible for larger code repositories \cite{Reimers2020}. 

The work reported on here is part of an industrial collaboration, where the company's agenda is to identify all clones in their very-large systems (100's of MLOC). T1 and T2 clones are currently detected using tools like CCFinder \cite{Kamiya2002654}) but their ability to find T3 and T4 clones is compromised by the scalability of CDTs more suited to detecting those clone classes. Consequently, recall of T3/T4 clones with scalability, was the primary motivation for the collaboration. Precision was an important, secondary, concern as they did not want to waste developer's time reviewing any false-positives suggested. 

Hence, in this paper we derive and evaluate a NN-based, T3/T4-oriented solution that avoids $O(n^2)$ pairwise comparisons and uses a NN based on CodeBERT/GraphCodeBERT. CodeBERT/GraphCodeBERT were selected as a NN basis, because these models seem to demonstrate state-of-the-art clone detection accuracy for pairwise clone comparison as compared to other novel NN models \cite{Guo2021}. But we fine-tune the resulting NN models with an algorithm based on contrastive loss towards a model that is able to distinguish between multiple code fragments. As with other NNs based on CodeBERT, the output of our model is an \textit{embedding}; that is a vector of numeric values that summarizes the input code fragment.

To avoid $O(n^2)$ comparisons between the embeddings for each of the $n$ code fragments, we use an efficient approximate $k$-nearest neighbour ($k$-NN) algorithm. This algorithm finds the $k$ nearest neighbours for each of the $n$ embeddings in $O(k N log(N))$ time. Given the set of $k$ nearest neighbours for each embedding, we can identify a ranked list of embeddings that are globally most similar (subject to the limitation that we find at most $k$ clones for each input fragment). This allows us to scale our clone detection approach to industrial-sized codebases with hundreds of millions of lines of code. 

We assess three fundamental parameters of the approach that can largely affect accuracy and efficiency: the (code) input length to the NN model, the inclusion of structural information, and the usage of an approximate K nearest neighbour algorithm (kANN). This leads to the first research question (RQ1): \textbf{How should such an approach be configured for optimal effectiveness/efficiency?} The second research question (RQ2) assesses \textbf{How such an approach compares to existing scalable CDTs?} The contributions are as follows:
\begin{itemize}[leftmargin=*]
    \item A proposed new solution to the problem of $O(n^2)$ pairwise comparisons in neural network-based clone detection, oriented towards T3/T4 clone detection. We use a BERT-based neural network to generate embeddings for each code fragment, and an efficient kANN to find similar embeddings.
	\item To do so, we propose contrastive loss based retraining to improve the ability of the neural network to generate embeddings that better distinguish between code fragments.
    \item We configure and evaluate our approach on two datasets representative of the clone scenarios present in industrial software at our collaborating company. These are 80 KLOC of C code and 424 KLOC of C++ code respectively. We also do so for the Java BigCloneBench benchmark.
    \item The evaluation performed suggests that our approach achieves significantly higher rates of recall than classical CDTs, such as SourcererCC and SAGA, while achieving faster execution times. Further, by using highly optimized off-the-shelf kNN software for graphics processing units (GPUs), it can perform the search significantly faster than almost all existing approaches, when GPUs are available.
    \item The industrially-representative C/C++ datasets are made available for other researchers in the field\cite{linkDataset}.
\end{itemize}

The paper is organized as follows: in Section~\ref{sec:related_work}, existing scalable CDTs and their limitations are discussed, followed by an introduction to state-of-the-art (SOTA) transformer-based ANN approaches. In Section~\ref{sec:approach}, the proposed approach is described and neural network fine-tuning details are provided. Section~\ref{sec:methodology} talks to the design of the empirical study used for configuring/evaluating SSCD and Section~\ref{sec:results} presents the evaluation results based on the three datasets employed. Later, in Section~\ref{sec:threats}, threats to validity are discussed. Section~\ref{sec:conclusions} concludes this paper, outlining future work.

\section{Background and Related Work}
\label{sec:related_work}

\begin{table}[!th]
\caption{Existing scalable CDTs}
\label{tbl:scalable_cdts}
\centering
\resizebox{\columnwidth}{!}{
\begin{tabular}{l|rccc}
\hline
\multicolumn{1}{c|}{Technique} & \multicolumn{1}{c}{\begin{tabular}[c]{@{}c@{}}Year of \\ publication\end{tabular}} & CDT type & \begin{tabular}[c]{@{}c@{}}Implementation\\ publicly available\end{tabular} & \begin{tabular}[c]{@{}c@{}}Seems to scale\\ to full BCB\end{tabular} \\ \hline
CCFinder                       & 2002                                                                               & token    & Y                                                                           & Y                                                                    \\
Deckard                        & 2007                                                                               & tree     & Y (2019 ver.)                                                                            &                                                                      \\
NiCAD                          & 2008                                                                               & text     & Y                                                                           &                                                                      \\
SourcererCC                    & 2016                                                                               & token    & Y                                                                           & Y                                                                    \\
CloneWorks                     & 2017                                                                               & token    & Y                                                                           & Y                                                                    \\
CCAligner                      & 2018                                                                               & token    &                                                                             &                                                                      \\
Oreo                           & 2018                                                                               & metrics  & Y                                                                           & Y                                                                    \\
SAGA                           & 2020                                                                               & token    &                                                                             & Y                                                                    \\ \hline
\end{tabular}
}
\end{table}

\subsection{Scalable Techniques for Clone Detection}
\label{subsec:rel_clones}
Clone detection is an active research area and, in the past twenty years, many CDTs have been proposed\cite{Rattan2013, Ain2019}. Fewer CDTs demonstrate evidence towards clone detection at scale, though. Table~\ref{tbl:scalable_cdts} summarizes well-known CDTs that have demonstrated the ability to scale to at least 10 MLOC in the literature (usually evaluated using a reduced version of the BigCloneBench with 13 MLOC (BCB13)). Some approaches have been shown to scale further to the entire BCB, which was originally about 250 MLOC\cite{Svajlenko2014a}, but this has increased over time. The table also captures the year of first publication, the type of CDT, and if the implementation is publicly available. As can be seen, the majority of these approaches are token-based (5/8), 5 have publicly available implementations, and 5 are reported as scaling to the entire BCB. 

Token-based CDTs transform source code into a stream of tokens where literals are usually substituted with common tokens to address T2 clone differences. Scalable token-based CDTs, presented in the Table~\ref{tbl:scalable_cdts}, can be divided into two groups. The first group, represented by CDTs like CCFinder\cite{Kamiya2002654} and SAGA\cite{Li2020} employ suffix trees/arrays for code-fragment and function-level \cite{Rattan2013} clone search. Here, preserving the order of tokens as they appear in source code is important for the search and both CCFinder and SAGA are very effective for T1/T2 clones, but begin to struggle with T3/T4 clones as gaps in the token sequences (representing the code) start to appear. To address this, CCFinder eliminates certain tokens from generated sequences, which allows for some simple T3 clone detection, and SAGA represents T3 clones by grouping T1/T2 clones together while allowing gaps \cite{Kamiya2002654, Li2020}. Notwithstanding this, the T3/T4 issues remain, as reflected by the experience of our industrial partner: CCFinder has low recall for their T3 clones. 

The other group, including SourcererCC and CloneWorks\cite{Sajnani2016, Svajlenko2017}, represent source code as bags-of-tokens (CCAligner - as a set of sequences of tokens \cite{Wang2018}) where their order is less important. They usually operate at function and file-level granularity. For these approaches, the bags-of-tokens representation paired with information retrieval (IR) algorithms allows for more accurate identification of T3/T4 clones. But IR algorithms seem to be less efficient than, for example, suffix trees, and both SourcererCC and CloneWorks use aggressive filtering to reduce the search space. In terms of scalability, CCAligner was not evaluated for more than 10 MLOC in the original paper\cite{Wang2018}. SourcererCC detects clones in 250 MLOC in 4.5 days\cite{Sajnani2016}. 

Oreo is the final approach that has been reported on for full BCB. It is a hybrid, IR-and-NN type approach that also uses software metrics. For T3/T4 clones, a trained ANN model is used to detect if all given candidate pairs are clones. But, to limit pairwise comparison, Oreo utilizes similarity heuristics based on metrics like size and call-similarity, to pre-filter candidates. While Oreo improves detection of T3/T4 clones, the filtering strategy still seems inefficient for large codebases, with Oreo performing six times slower than SourcererCC\cite{Saini2018}. 

In terms of the other approaches in Table~\ref{tbl:scalable_cdts}, that have not been reported as working on the full BCB, Deckard is an Abstract Syntax Tree-based CDT suitable for block and function level clone detection \cite{Jiang2007}. The reason it has not been reported on for the full BCB is that it returns 400GB of clone pairs for that dataset and the evaluation tool (BigCloneEval) fails on such a large amount of data. NiCAD is a text-based CDT that can detect block/function level clones \cite{Roy2008172}. To search for clones, it uses the Longest Common Subsequence algorithm and that becomes inefficient for large codebases.

In contrast, SSCD, can be text or tree-based and, unlike the CDTs described in this section, looks at the context of words in the code, addressing synonyms and polysemy. Also, the context is large and bidirectional allowing for better T3/T4 clone detection. It does so by adopting an efficient BERT-based architecture \cite{vaswani2017attention}. This architecture is now described.

\subsection{BERT-based Architectures for Clone Detection}
\label{subsec:rel_bert}
A \textit{pre-trained model} is an ANN model that has been trained towards a generic set of tasks (e.g. a set of natural language tasks). In a transfer learning approach \cite{pan2009survey}, existing learnt knowledge via the pre-training can be re-used for other tasks, based on subsequent fine-tuning training; the weights in the pre-trained model are not finalized, allowing for further learning. This results in a \textit{fine-tuned model}. 

\begin{figure*}[htbp]
\centerline{\includegraphics[width=0.85\textwidth]{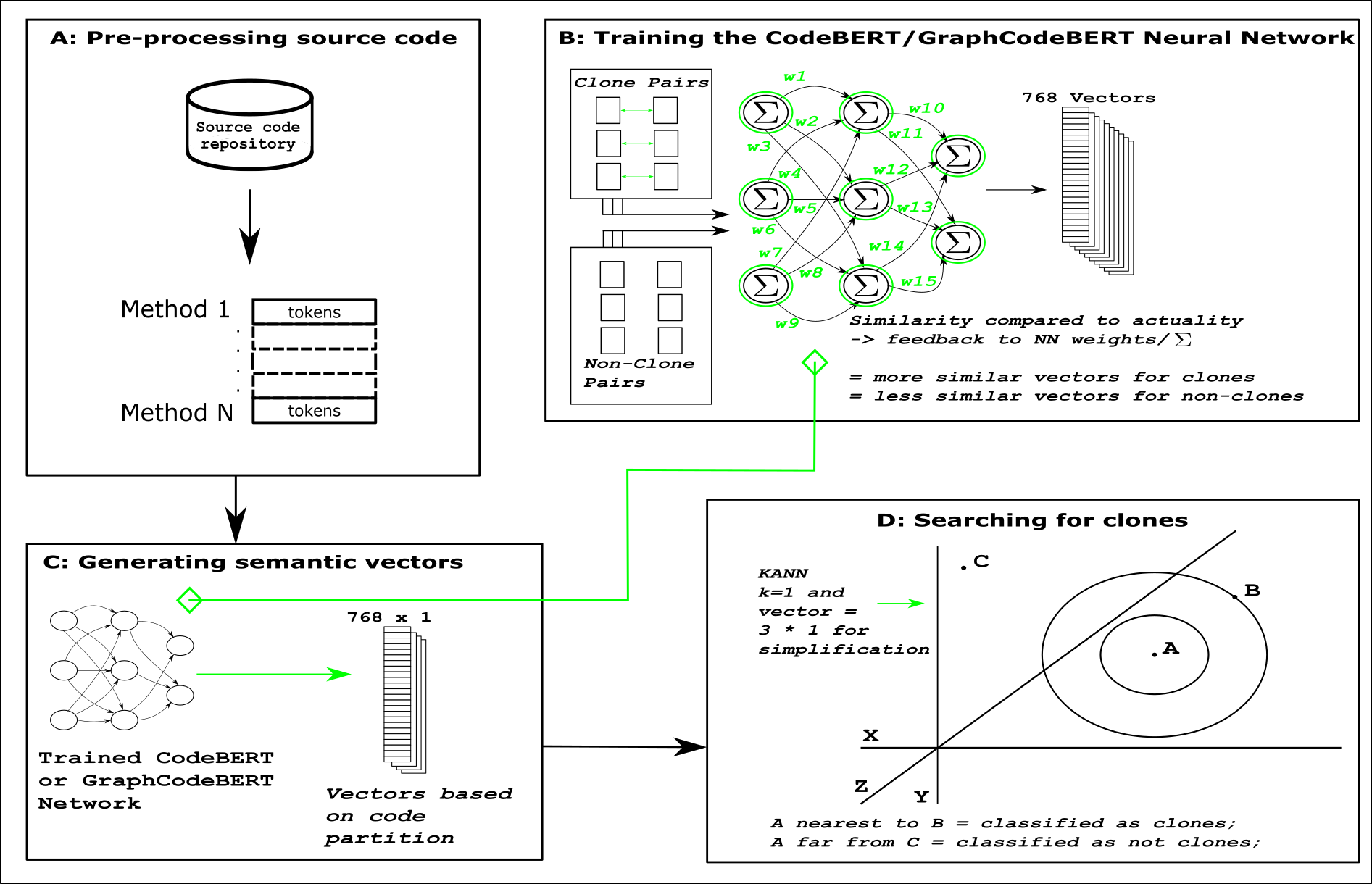}}
\caption{The architecture of SSCD}
\label{fig1:sscd-stage-01}
\end{figure*}

For example, a generic pre-trained BERT natural language model can be further fine-tuned towards a question-answering task, where BERT is a pre-trained ANN model utilizing the Transformer architecture \cite{vaswani2017attention} that has achieved SOTA results recently for a number of natural language tasks\cite{Devlin2018}. The architecture is multi-layer with attention heads that allows for bidirectional capture of words' context in a sequence (see Vaswani et al. \cite{vaswani2017attention}). The BERT authors experimented with two architectures: \textbf{base} that has 12 layers, 768 hidden size (a vector representing context), and 12 attention heads and \textbf{large} that has 24 layers, 1024 hidden size, and 16 attention heads. 

The success of this architecture prompted its application to other areas with natural and structured text. In this vein two BERT-derivative pre-trained models were produced for application to source code tasks: CodeBERT and GraphCodeBERT \cite{Feng2020, Guo2021}. These were subsequently fine-tuned for clone detection using a subset of BigCloneBench clones and they reported SOTA results for clone detection/prediction (with GraphCodeBERT outperforming CodeBERT) \cite{Feng2020,Guo2021}. 

\begin{figure}[htbp]
\centering
\includegraphics[width=0.85\columnwidth]{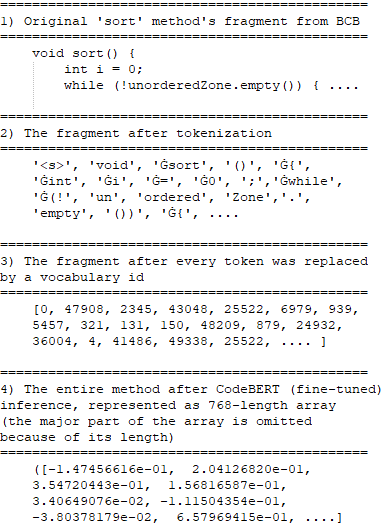}
\caption{CodeBERT/GraphCodeBERT tokenization and vectorization}
\label{fig:example}
\end{figure}

The architecture of these two models is similar (utilizing RoBERTa \cite{liu2019roberta}), but their inputs are different: CodeBERT is trained and fine-tuned using textual information only --- a vector of 512 (maximum size) textual, source-code tokens. GraphCodeBERT, in contrast, is trained and fine-tuned using textual and structural information. This is a vector of 640 (maximum size) token ids, 512 of which contain textual information and 128 of which contain structural information. It also leverages a vector of 640 position ids, and a corresponding attention matrix\cite{Guo2021}. This difference has significant effects on training and inference time, with GraphCodeBERT being more resource-intensive. Also, the tokenization employed for both is different to BERT, with CodeBERT/GraphCodeBERT using tokenization similar to the GPT-2 model and Byte-Pair Encoding (see Fig. \ref{fig:example}) \cite{sennrich2015neural}. During tokenization words that aren't first in a sentence are preceded with whitespaces. Subsequently the tokens are mapped to their vocabulary identifiers as the basis for deriving the code's vector representation. 

The fine-tuning in \cite{Feng2020,Guo2021}  was done for pairwise clone prediction (labelling each two pieces of code as clones or not). As noted, this approach is costly to scale to larger codebases. 

\section{The Proposed Approach: SSCD}
\label{sec:approach}
The idea behind SSCD is to find similar code fragments using their numerical representations. These representations are generated by ANNs and their similarity is established using a kNN approach. An overview of SSCD is shown in Fig.~\ref{fig1:sscd-stage-01}.

\subsection{Training CodeBERT/GraphCodeBERT for Scalable Clone Detection}
\label{subsec:fine-tune}
The original CodeBERT and GraphCodeBERT models are inefficient for clone detection because they compare every pair of source code fragments individually\cite{Feng2020, Guo2021}. This does not scale well and so, in this work:
\begin{itemize}
    \item The architecture of pre-trained CodeBERT and GraphCodeBERT models was modified to add a pooling layer that returns a 768-dimensional vector representation of code pieces for comparison, as shown in Fig. \ref{fig1:sscd-stage-01} (B). An averaging function similar to Reimers et al. \cite{Reimers2020} was used.
    \item Fine-tuning for clone detection is achieved using a contrastive loss function during training, allowing better discrimination between features of inputs and better accuracy as compared to a cross-entropy function\cite{lian2018speech}. 
\end{itemize}
For CodeBERT-based SSCD, fine-tuning was using 433,000 Java clone pairs from the CodeBERT/GraphCodeBERT clone datasets (derived, in turn, from BigCloneBench) \cite{Feng2020,Guo2021}. The hyper-parameters were: 1 epoch, 16-batch size for training, 32-batch size for validation, and AdamW optimizer. The fine-tuning was performed on Google Colab with K80 GPU. For the GraphCodeBERT incarnation, fine-tuning was performed using the same Java dataset. The hyper-parameters were: 1 epoch, 8-batch size for training, 8-batch size for validation, and AdamW optimizer. Due to the resource-intensive fine-tuning process, the work was performed on an Amazon Sagemaker multi-GPU machine with 4 GPUs (64 GB GPU memory total). Interestingly, loss and accuracy started to decline for this model after 99,000 clone pairs, possibly due to overfitting. Hence, while three models were saved/used in the evaluation, only the fine-tuned CodeBERT (CBf), and the fine-tuned GraphCodeBERT saved after 99K clone pairs (GCBf99) are presented here. The fine-tuned GraphCodeBERT, after 433K clone pairs, did not perform as well.

\subsection{Clone Detection with SSCD}
\textbf{Pre-processing source code}. The input to SSCD is a source code repository. In this step, the source code is divided into chunks of a certain granularity using a parser; we use method-level granularity, but there are no architectural limitations to this. It is also possible to remove comments from the method bodies and to specify a minimum LOC threshold, 6 LOC being commonly specified in the literature\cite{Sajnani2016}. After this, every method is represented by a list of tokens (see Fig.~\ref{fig1:sscd-stage-01} (A)). 

\textbf{Generating semantic vectors.}
Tokens, representing source code methods, are used to generate semantic numerical vectors based on either the CBf, or GCBf99 fine-tuned models, as shown in Fig.~\ref{fig1:sscd-stage-01}, part C. For CBf, only textual information is used to generate vectors. For GCBf99 additionally structural information is added and the input is expanded into three vectors, as described in Section \ref{subsec:rel_bert}. The output of the models is always a 768-dimensional vector, regardless of the input size and Fig.~\ref{fig:example} illustrates these steps using CBf. First, the ``sort'' method (1) is tokenized into a sequence of tokens (2), where the special token ``G'' represents white-space and special tokens $<s>$/$</s>$ mark the start and end of this sequence. Next, all tokens are replaced with vocabulary ids (3) and these are used as an input to the CBf neural network. The output (4) is a 768-dimensional vector representing the ``sort'' function.

\textbf{Searching for clones.} The numerical vectors produced are then used by the search module, illustrated in part D of Fig.~\ref{fig1:sscd-stage-01} for a simplified vector of three dimensions (X, Y and Z). Here, SSCD allows for either an exact brute-force kNN search or a kANN-based (K approximate nearest neighbour) search, as exact kNN approaches can be inefficient\cite{Malkov2018}. Several approximation techniques such as locality sensitive hashing, product quantization with GPUs \cite{Johnson2019}, and using ``navigable small worlds'' \cite{Malkov2018} exist. In this work, ``navigable small worlds'' was selected for kANN search (implemented using ``hnswlib'' as we empirically found it more efficient than ``faiss'' with the same parameters) because of its logarithmic $O(log(N)$ search complexity and $O(N log (N))$ construction complexity. The exact search performs pairwise comparison between all generated vectors and orders them based on their cosine similarity. The kANN search utilizes the kANN model for efficiency, where cosine similarity is again used to calculate distances and order the results.

\textbf{Displaying the results.} For each source code method, similar methods are placed in ranked order. Two parameters control the size of this ranked list: similarity threshold and topN. The former is a cosine similarity threshold and the latter controls the number of returned results. All these lists are merged into one ordered list, ranked on similarity.

\textbf{Implementation details.} SSCD is a command-line application that uses the following libraries: ``Tree\_sitter''\footnote{\url{https://github.com/tree-sitter/tree-sitter}} is used to parse source code and to extract textual/structural information. ``hnswlib''\footnote{\url{https://github.com/nmslib/hnswlib}} and ``faiss''\footnote{\url{https://faiss.ai/}} are used for kNN/kANN implementation\cite{Malkov2018}. Finally, ``transformers''\footnote{\url{https://huggingface.co/transformers/}} and ``sbert''\footnote{\url{https://www.sbert.net/}} libraries are used to support neural network operations.

\section{Design of the Empirical Study}
\label{sec:methodology}
\subsection{Empirical Objectives}
\label{subsec:E_Objectives}
The empirical work reported here aims to identify an efficient/effective configuration for SSCD and to rate its effectiveness/efficiency with respect to other CDTs capable of working on large/industrial systems. As such, the evaluation breaks down into the following objectives:

\begin{itemize}
    \item Determining the best-performing configuration of SSCD on realistic datasets, in terms of a) input size (128 or 512 tokens), b) employing textual or textual-and-structural input and c) employing kANN versus exact kNN search. 'Realistic' here refers to code representative of clones in industrial codebases.    
    \item  Assessing the recall-effectiveness and efficiency of SSCD compared to other CDTs capable of working on very large-scale datasets. As it is impossible to determine recall using industrial datasets where all the clones are not known in advance, the dataset for this evaluation was full BCB (now approximately 320 MLOC). Precision, as a secondary concern, was also assessed.
\end{itemize}

\subsection{Evaluation Design}
\label{subsec:methodology}
To configure SSCD and to assess its effectiveness/efficiency, four clone datasets were used (see Table~\ref{tbl:datasets}), at method level.

\begin{table}[!th]
\caption{Clone datasets}
\label{tbl:datasets}
\centering
\resizebox{\columnwidth}{!}{\begin{tabular}{l|rrlrr}
\hline
Dataset     & Total LOC                       & \# Methods                                                                   & Language & \# Files                    & \begin{tabular}[c]{@{}r@{}}\# Clone \\ pairs\end{tabular} \\ \hline
Company-C   & 80,190                          & 1,714                                                                        & C        & 160                         & 80                                                        \\
Company-C++ & 424,626                         & 18,500                                                                       & C++      & 200                         & 100                                                       \\
BCB13       & 13,357,013                      & 854,050                                                                      & Java     & 55500                       & 8,375,313                                                 \\
BCB         & \multicolumn{1}{l}{320,217,373} & \begin{tabular}[c]{@{}r@{}}4,882,375\\ (\textgreater{}= 10 LOC)\end{tabular} & Java     & \multicolumn{1}{l}{2876219} & \multicolumn{1}{l}{8,375,313}                             \\ \hline
\end{tabular}
}
\end{table}

\textbf{Clone datasets.} Our partner company is interested in locating clones in C/C++ code as well as Java. However, a literature review suggests that only the Murakami et al. benchmark is openly available for those languages \cite{Murakami2014}. On closer inspection, the company's experts decided that Murakami's benchmark did not reflect their code and they manually designed more representative datasets of clones, consisting of both (C/C++) source code and known clones embedded in that code, spanning the four clone types. These datasets are available online \cite{linkDataset} to provide transparency, and as a resource for other researchers. In this instance, they were used to determine SSCD's best-performing configuration and to compare it to a currently available CDT on company-relevant code.

BigCloneBench \cite{Svajlenko2014a} is a frequently used Java clone benchmark. It contains a database of known clone pairs for the IJADataset: a compilation codebase of Java open source projects. In this evaluation, two versions of the BigCloneBench benchmark were used: the reduced version of BCB (in terms of LOC, not clone pairs) containing 13 MLOC was used to further refine SSCD's configuration, and the entire BCB, containing 320 MLOC (according to the ``cloc'' tool) was used to evaluate the effectiveness and efficiency of the approach.

\textbf{CDT Selection.} Selection of a CDT for comparison was based on public availability, ability to scale well to larger codebases and ability to locate T3 and T4 clones (see Table ~\ref{tbl:scalable_cdts}), as per the company's requirements. Based on these criteria, SourcererCC was selected for comparison. CCFinder, was already tried by the Company and was found ineffective/inefficient. CloneWorks seems functionally similar to SourcererCC (see Section~\ref{sec:related_work}) and Oreo, despite its good recall, was reported as prohibitively slower than SourcererCC \cite{Saini2018}.

\textbf{CDT parameters.} Table~\ref{tbl:params} summarizes the major parameters specific to SSCD and to SourcererCC that were used in this work. ``minLOC'' and ``minTokens'' are cutoff values in SSCD and SourcererCC respectively: methods below this length are excluded. It was assumed that 1 LOC corresponds to 5 tokens, which seems to be acceptable practice \cite{Sajnani2016}. 0 and 6 LOC (30 tokens) were specified for the C/C++ datasets, the former solely to provide an efficiency stress-test for SSCD. For both BCB benchmarks 10 LOC (50 tokens) was used. 

Two fine-tuned models, CodeBERT (CBf) and fine-tuned GraphCodeBERT (GCBf99), were assessed for SSCD (see Section~\ref{subsec:fine-tune}) to check on the importance of including structural information. Each of these models can take either 128 or 512 long code pieces. The search type of SSCD can be set to either exact or hnsw (kANN) (see Section~\ref{sec:approach}). HNSW, in turn, has its own three major parameters: M (the number of neighbours in the graph), EFC (the depth of a graph traversal during construction), and EFS (the depth of a graph traversal during search). The latter were set to 32, 200, and 120 respectively for better accuracy (at the expense of time). TopN limits the number of candidates per code fragment returned by SSCD. Finally, similarity in SSCD is a cosine similarity threshold: code pieces below that similarity are ignored. In SourcererCC, similarity seems to refer to textual similarity between code fragments \cite{Sajnani2016} and again, code pieces below that threshold are ignored.

\begin{table}[!th]
\caption{CDT parameters}
\label{tbl:params}
\centering
\resizebox{0.75\columnwidth}{!}{

\begin{tabular}{l|l|ll}
\hline
CDT                          & Symbol & Description & Values used        \\ \hline
\multirow{6}{*}{SSCD}        & $\alpha$       & minLOC      & 0 / 6 / 10         \\
                             & $\beta$       & NN model    & CBf / GCBf99       \\
                             & $\gamma$       & codeLength  & 128 / 512          \\
                             & $\delta$       & searchType  & exact / hnsw        \\
                             & $\epsilon$       & topN        & 0 / 1 / 100 \\
                             & $\sigma$       & similarity  & 0 / 0.95       \\ \hline
\multirow{2}{*}{SourcererCC} & $\eta$       & minTokens   & 30 / 50        \\
                             & $\theta$       & similarity  & 50\% / 70\%       \\ \hline
\end{tabular}}
\end{table}

\textbf{Machine configurations.} Two machines were used in this work for SSCD's configuration/evaluation. A CPU-based \textbf{M1} with i7-10875H 2.3GHz 8 core CPU, 32 GB RAM, and 1TB SSD and a GPU-based \textbf{L1} with i7-6700K 4GHz 8 core CPU, 6x8GB GeForce GTX 1070, 16 GB RAM, and 1TB SSD.

\textbf{Metrics.} In clone detection recall and precision, in combination, are used to assess the effectiveness of CDTs\cite{Rattan2013}. The former tells how many known clones were identified: if 50 of 100 known clones were located by a CDT, then its recall is 50\%. The latter (precision) informs on how accurate it is: if a CDT returns 100 clone candidates and 30 of these are correct, then its precision is 30\%. In industrial practice both are important in combination, as low precision demands developers discount many false positives and low recall suggests a high miss rate. Hence for the studies on the C/C++ datasets we report precision and recall, but we also report an F-score: multiplication of precision and recall divided by their sum and multiplied by 2. This gives a normalized, unified measure of both concerns. Finally, because SSCD returns ranked lists, mean reciprocal rank (MRR) is also calculated. This metric assesses how close to the top of the list the first correct result is, averaged over all searches\cite{chochlov2017historical}.

Such an approach is not suitable for the BCB datasets as not all clones in the BCB are explicitly identified. So, while a recall-type measure can report on the proportion of clones found by the CDT that are explicitly-identified in the dataset, a proxy for precision is more difficult to capture. In this work a random sample of (400) clones identified by the CDT approach was obtained and the code segments manually inspected by two researchers to determine if each identified 'clone' was a true-positive or a false positive. Reliability of the manual analysis is assessed using the Cohen Kappa \cite{mchugh2012interrater, kappaLink} and, for those where the human coders were in agreement, the proportion of true positives and false positives were employed to define a proxy for precision. When precision is calculated for CDTs, this sampling of 400 candidates seems standard practice in the field \cite{Li2020,Sajnani2016,Saini2018}. Reliability of this analysis was 'substantial' with both coders agreeing on 86.75\% of cases (a Kappa of 0.62). The reliability data is presented in Table~\ref{tbl:rev_rel}.   

\begin{table}[!th]
\caption{Reviewer inter-rater agreement}
\label{tbl:rev_rel}
\centering
\resizebox{0.75\columnwidth}{!}{
\begin{tabular}{l|rr}
\hline
\begin{tabular}[c]{@{}l@{}}\backslashbox{Reviewer 2}{Reviewer 1}\end{tabular} & Clone & Non-clone \\ \hline
Clone                                                           & 285   & 32        \\
Non-clone                                                       & 21    & 62        \\ \hline
\end{tabular}
}
\end{table}

To assess the efficiency of CDTs, execution time was measured. For SSCD, this was divided into inference time (the time that it takes for a neural network to produce numerical vectors from all its code input) and search time (the time it takes for SSCD's search engine to find clone candidates). For SourcererCC, this was calculated as the total time between when the CDT was launched and when it completed.

\section{Configuration/Evaluation Results}
\label{sec:results}
\subsection{(RQ1) How should such an approach be configured for effectiveness/efficiency?}

\begin{table*}[htbp]
\caption{Results of using SSCD configurations with Company's datasets ($\alpha=0, \epsilon=10, \sigma=0$)}
\label{tbl:results_company}
\centering
\resizebox{0.8\textwidth}{!}{
\begin{tabular}{l|rrrrr|rrrrr}
\hline
\multicolumn{1}{c|}{\multirow{2}{*}{\begin{tabular}[c]{@{}c@{}}SSCD \\ configurations\\ ($\gamma, \delta, \beta$)\end{tabular}}} & \multicolumn{5}{c|}{Company-C}                                                                                                                                                                                                                                                                        & \multicolumn{5}{c}{Company-C++}                                                                                                                                                                                                                                                                       \\ \cline{2-11} 
\multicolumn{1}{c|}{}                                                                                            & \begin{tabular}[c]{@{}r@{}}MRR \\ (\%)\end{tabular} & \begin{tabular}[c]{@{}r@{}}Recall \\ (\%)\end{tabular} & \begin{tabular}[c]{@{}r@{}}Precision \\ (\%)\end{tabular} & \begin{tabular}[c]{@{}r@{}}Inference\\ Total (s)\end{tabular} & \begin{tabular}[c]{@{}r@{}}Search\\ Total (s)\end{tabular} & \begin{tabular}[c]{@{}r@{}}MRR \\ (\%)\end{tabular} & \begin{tabular}[c]{@{}r@{}}Recall \\ (\%)\end{tabular} & \begin{tabular}[c]{@{}r@{}}Precision \\ (\%)\end{tabular} & \begin{tabular}[c]{@{}r@{}}Inference\\ Total (s)\end{tabular} & \begin{tabular}[c]{@{}r@{}}Search\\ Total (s)\end{tabular} \\ \hline
128 exact CBf                                                                                                    & 78.18                                               & 86.08                                                  & 2.31                                                      & 84.16                                                         & 0.04                                                       & 85.11                                               & 88.76                                                  & 0.35                                                      & 835.17                                                        & 0.06                                                       \\
512 exact CBf                                                                                                    & 79.16                                               & 86.08                                                  & 2.31                                                      & 408.23                                                        & 0.04                                                       & 89.33                                               & 89.89                                                  & 0.36                                                      & 2764.16                                                       & 0.06                                                       \\
128 hnsw CBf                                                                                                     & 78.18                                               & 86.08                                                  & 2.31                                                      & 85.62                                                         & 0.03                                                       & 85.11                                               & 88.76                                                  & 0.35                                                      & 848.63                                                        & 0.22                                                       \\
512 hnsw CBf                                                                                                     & 79.16                                               & 86.08                                                  & 2.31                                                      & 410.06                                                        & 0.03                                                       & 89.33                                               & 89.89                                                  & 0.36                                                      & 2864.82                                                       & 0.23                                                       \\
128 exact GCBf99                                                                                                 & 82.7                                                & 86.08                                                  & 2.31                                                      & 252.48                                                        & 0.06                                                       & 88.76                                               & 89.89                                                  & 0.36                                                      & 2197.51                                                       & 0.07                                                       \\
512 exact GCBf99                                                                                                 & 83.97                                               & 86.08                                                  & 2.31                                                      & 772.21                                                        & 0.07                                                       & 88.2                                                & 89.89                                                  & 0.36                                                      & 6403.28                                                       & 0.07                                                       \\
128 hnsw GCBf99                                                                                                  & 82.7                                                & 86.08                                                  & 2.31                                                      & 253.05                                                        & 0.04                                                       & 88.76                                               & 89.89                                                  & 0.36                                                      & 2232.47                                                       & 0.24                                                       \\
512 hnsw GCBf99                                                                                                  & 83.97                                               & 86.08                                                  & 2.31                                                      & 634.75                                                        & 0.03                                                       & 88.2                                                & 89.89                                                  & 0.36                                                      & 6421.25                                                       & 0.24                                                       \\ \hline
\end{tabular}}
\end{table*}

To answer this question, SSCD was applied on the Company's test datasets and the results are shown in Table~\ref{tbl:results_company}. Here topN was set to 10, there was no limit on similarity, the LOC cut-off was set artificially low to zero, and embedding inference/search was done on a CPU (machine M1, see Section~\ref{sec:methodology}, ``Machine configurations''). Such choice of parameters and hardware was selected to stress test and to highlight efficiency differences. Three configuration parameters were assessed: the input length, the inclusion of structural information in NN (CBf vs GCBf99), and the usage of kANN. As can be seen:
\begin{itemize}
    \item The 512 code-length configurations gave higher (better) MRR in 6/8 cases. The longer code length cut-off means that more textual information was processed by the neural network, thus increasing detection accuracy.
    \item Longer code lengths and inclusion of structural information (in GCBf99) also resulted in substantially increased inference time as compared to shorter code length and a text-only model (CBf). This is due to an architectural characteristic of BERT-based models, where longer token streams take more time to process (see Section~\ref{subsec:rel_bert}), and the increased complexity of the GCB inputs.
    \item In terms of effectiveness, the precision and recall differences across all configurations were minimal. While recall was high, precision was low, as is to be expected with such a choice of parameters.
\end{itemize}

To summarize, although longer input lengths (512) and the inclusion of structural information in GCBf99 resulted in slightly increased effectiveness, the inference time increased several-fold. For example, for 128 exact CBf versus 512 exact GCBf99 the inference time increased almost 9 times for the Company-C dataset and over 7 times for the Company-C++ dataset with little-to-no improvements in recall/precision. The search time efficiency results for kANN are mixed: the kANN search time was better than exact search time in 4/4 cases for the Company-C dataset, but worse in all 4/4 C++ cases.

\begin{table}[htbp]
\renewcommand{\arraystretch}{1.5}
\caption{Results of using SSCD with BCB13}
\label{tbl:rq1_bcb}
\centering
\resizebox{\columnwidth}{!}{

\begin{tabular}{l|cccccc|r}
\hline
\multirow{2}{*}{Configuration} & \multicolumn{6}{c|}{Recall (\%)}     & \multirow{2}{*}{\begin{tabular}[c]{@{}r@{}}Search time\end{tabular}} \\
                               & T1  & T2 & VST3 & ST3 & MT3 & WT3/T4 &                                                                                                     \\ \hline
SSCD ($\delta=exact$, CPU)              & 100 & 97 & 96   & 81  & 28  & 2      & 23m 00s                                                                                             \\
SSCD ($\delta=hnsw$, CPU)               & 100 & 96 & 97   & 81  & 28  & 2      & 3m 23s                                                                                              \\
SSCD ($\delta=exact$, 1 GPU)            & 100 & 97 & 96   & 81  & 28  & 2      & 2m 08s                                                                                              \\
SSCD ($\delta=exact$, 5 GPU)            & 100 & 96 & 97   & 81  & 28  & 2      & 59s                                                                                                 \\ \hline
\end{tabular}
}
\end{table}

To further assess the search algorithm, towards the search time, SSCD was used with a larger BCB13 clone benchmark on machine L1. Here, the parameter $\alpha=10$ (minLOC) was employed as this has been used by the community before in BCB evaluations \cite{Svajlenko2014a}. $\gamma=128$ and $\beta=CBf$, based on the above assessment, favouring faster inference time over marginal advances in effectiveness. $\epsilon=100$ and $\sigma=0.95$ (topN and similarity), as sampling of the BCB13 benchmark suggested large clone classes compared to the company's datasets. The results are shown in Table~\ref{tbl:rq1_bcb}. As can be seen, the differences in recall are marginal but, with a CPU-based configuration, HNSW kANN is almost 6.5 times faster than exact search. However, GPU-based exact searches are faster than CPU-based kANN. Multi-GPU improve this further: the search index was replicated over 5 GPUs for fast search. GPU-based HNSW is currently unrealized in the \textbf{faiss} library, but existing research suggests the possibility of GPU-HNSW \cite{zhao2020song}.

To answer RQ1, shorter input lengths and a textual-only CodeBERT model seems to be preferable for scalable clone detection with SSCD due to significant inference time advantages. Longer-length, textual-structural modelling seems to only slightly improve clone detection effectiveness with SSCD. On a CPU-only machine, kANN HNSW significantly outperforms exact search (retaining effectiveness) and thus should be used. On a GPU-based machine, exact search seems to work faster than HNSW (with GPU-parallelism improving exact search further). It remains to be seen if HNSW GPU-based algorithms can be advantageous in such a setup.  

\subsection{(RQ2) How does such an approach compares to existing scalable CDTs?}
To answer this question, SSCD was compared to existing CDTs using both the company's datasets (on M1) and the BCB full dataset (on L1), in terms of effectiveness and efficiency. 

\subsubsection{Results for the Company's Datasets}
\begin{table}[!th]
\renewcommand{\arraystretch}{1.5}
\caption{Comparison of SSCD and SourcererCC on Company's datasets}
\label{tbl:sscd_vs_sourcerer_company}
\centering
\resizebox{\columnwidth}{!}{
\begin{tabular}{l|crr|crr}
\hline
\multirow{2}{*}{Approach}      & \multicolumn{3}{c|}{Company-C}                                                                                                                                                                                     & \multicolumn{3}{c}{Company-C++}                                                                                                                                                                                   \\ \cline{2-7} 
                               & \begin{tabular}[c]{@{}c@{}}Recall\\ (\%)\end{tabular} & \multicolumn{1}{c}{\begin{tabular}[c]{@{}c@{}}Precision\\ (\%)\end{tabular}} & \multicolumn{1}{c|}{\begin{tabular}[c]{@{}c@{}}F-score\\ (\%)\end{tabular}} & \begin{tabular}[c]{@{}c@{}}Recall\\ (\%)\end{tabular} & \multicolumn{1}{c}{\begin{tabular}[c]{@{}c@{}}Precision\\ (\%)\end{tabular}} & \multicolumn{1}{c}{\begin{tabular}[c]{@{}c@{}}F-score\\ (\%)\end{tabular}} \\ \hline
SourcererCC ($\theta=50\%$)                   & \multicolumn{1}{r}{86.08}                             & 14.81                                                                        & 25.27                                                                       & \multicolumn{1}{r}{85.39}                             & 2.07                                                                         & 4.04                                                                       \\
SSCD ($\epsilon=1, \sigma=0.95$) & \multicolumn{1}{r}{60.76}                             & 84.21                                                                        & 70.59                                                                       & \multicolumn{1}{r}{79.78}                             & 88.75                                                                        & 84.03                                                                      \\ \hline
\end{tabular}
}
\end{table}

SSCD was compared to SourcererCC, parameterized with 50\% textual similarity ($\theta=50\%$) and using a more realistic 6 LOC/30-token-minimum cut-off size in terms of code granularity ($\eta=30, \alpha=6$) as per the company's requirements. The 50\% similarity for SourcererCC was selected to include more T3/T4 clones, again as per company's requirements. For SSCD, $\epsilon=1, \sigma=0.95$ were selected as per the company's demand for a very accurate search and their awareness that there were limited clone classes in the datasets. (There are no alternatives for these two SSCD parameters in SourcererCC). The rest of SSCD's parameters follow the best configuration in terms of efficiency (see RQ1): 128-length CBf with exact search. As can be seen from the results in Table~\ref{tbl:sscd_vs_sourcerer_company}, SSCD achieved a significantly higher F-score on both datasets: 3 times better on the C dataset and 20 times better on the C++ dataset. It should be noted that, while this is a better score in terms of CDTs generally, it does not tally completely with our industrial collaborator's requirements. It did mean that developers were faced with substantially fewer false-positives returned by SSCD, but recall also decreased, meaning that a less complete set of clones was returned by the technique. 

The average total execution time for SourcererCC was 4 minutes and 36 seconds (SourcererCC re-creates a search index for every run and that adds significantly to the average execution time), and 1.5 seconds/10 seconds for SSCD on average for C/C++ datasets respectively.

\subsubsection{Evaluating SSCD recall with full BCB}
\label{subsubsec:full_bcb_eval}

When comparing CDTs for effectiveness, common practice is to either do an actual comparison using the same hardware settings or to compare against existing results from the literature \cite{Saini2018, Sajnani2016, Li2020}. As regards actively evaluating against other CDTs capable of performing at this scale (see Section~\ref{sec:related_work}), SourcererCC was selected for recall-based comparison (see Section~\ref{sec:methodology}, ``CDT selection'' for reasoning).

Of the two other scalable approaches, the authors of SAGA claim that the recall they report for BCB13 holds for the entire BCB and so these figures were included for comparison \cite{Li2020}, even though finding clones in the full BigCloneBench would seem more difficult, because of its large codebase (see Table~\ref{tbl:datasets}) and the 'noise' that that brings. In contrast, the authors of Oreo do not explicitly mention if recall was calculated for the entire BCB \cite{Saini2018} and so it was excluded from our analysis. 

For evaluation, the minimum number of LOC for SSCD was set to 10 ($\alpha=10$) (a standard for evaluations with BCB \cite{Sajnani2016}). For SourcererCC, the equivalent minimum number of tokens was set to 50 ($\eta=50$) (see Section~\ref{sec:methodology}, ``CDT parameters''). For SSCD, CBf with 128 token length and exact search was used as a result of the RQ1 evaluation, where smaller input lengths and exclusion of structural information yield significantly better efficiency at the expense of only marginal effectiveness. The BigCloneEval \cite{Svajlenko2017a} was used for recall calculation and it was configured with the following parameters: 70\% code overlap acceptance for detected clones and a minimum of 10 LOC for clones in the BCB database.

\begin{table}[!th]
\renewcommand{\arraystretch}{1.5}
\caption{Recall evaluation with SSCD and SourcererCC on the entire BCB}
\label{tbl:recall_bcb_full}
\centering
\resizebox{\columnwidth}{!}{
\begin{tabular}{l|crrrrrr}
\hline
\multirow{2}{*}{CDT}                                                        & \multirow{2}{*}{\begin{tabular}[c]{@{}c@{}}Data \\ source\end{tabular}} & \multicolumn{6}{c}{Recall (\%)}       \\ \cline{3-8} 
                                                                            &                                                                         & T1  & T2  & VST3 & ST3 & MT3 & WT3/T4 \\ \hline
SSCD ($\epsilon=100, \sigma=0.95$)                                                     & A                                                                       & 100 & 97  & 96   & 80  & 27  & 1      \\ \hline
SourcererCC ($\theta=70\%$)                                                        & A                                                                       & 85  & 94  & 52   & 52  & 2   & 0      \\ \hline
\begin{tabular}[c]{@{}l@{}}SAGA (best config\\ sim=60, lmc=40)\end{tabular} & L                                                                       & 100 & 100 & 95   & 60  & 10  & 0      \\ \hline
\end{tabular}}
\end{table}

The results of this recall evaluation are shown in Table~\ref{tbl:recall_bcb_full}. In the table, the ``Data source'' column shows the source of results. The letter ``A'' means that an actual evaluation was conducted (on L1 machine), whereas the letter ``L'' means that the results come from existing literature (SAGA).

SSCD outperformed SourcererCC for all clone types. It should be noted that SourcererCC has been evaluated previously using BCB13 and showed better recall: for example, in their original paper Sajnani et al. \cite{Sajnani2016} have reported 100\% T1, 98\% T2, 93\% VST3, 61\% ST3, 5\% MT3, and 0\% WT3/T4. But it is likely that these recall results do not hold for the entire BCB due to the larger codebase. It should also be noted that, when executing SourcererCC, a small number of parsing warnings were observed. This potentially affected the clone detection reported in this study slightly.

Compared to SAGA, SSCD shows the same recall for T1 clones and marginally worse recall for T2 clones. However, for all the other clone types (VST3, ST3, MT3, and WT3/T4), SSCD had better recall than SAGA. Here, $\epsilon=100, \sigma=0.95$ was used again due to BCB containing extremely large clone classes. Overall, SSCD demonstrates an ability to effectively find clones in large codebases. It also appears to outperform the existing SOTA scalable CDTs for at least 4/6 clones types.

\subsubsection{Evaluating SSCD precision with BCB}
To evaluate precision, a sample of 400 clone candidates was randomly selected from SSCD's output and then manually inspected by the paper's authors to identify clones; a common practice in the CDT literature \cite{Li2020,Sajnani2016,Saini2018}. When evaluating a clone candidate pair, it was examined to assess if the pair exercised the same functionality and/or if they were textually similar.

Over the 347 clone candidates where agreement was achieved, 285 were marked as clones. This results in 82.13\% precision. Given the requirements of our industrial collaborator (prioritizing high recall), there is an argument for reporting all pairs (both agreed \textbf{and} disagreed on), where disagreements are considered as potential clones (that should be inspected by the developer). In that case ``(optimistic) precision'' is 338/400 or 84.5\%. And, to be even-handed in our analysis, we also report '(pessimistic) precision' where disagreements are considered as non-clones, in which case precision is 71.25\%.

Precision results reported by other techniques vary. (Here we largely rely on the precision figures compiled by the SAGA authors \cite{Li2020} for BCB13, so again the noise introduced by scaling up to the full BCB puts our precision at a comparative disadvantage.) According to these authors, SAGA, CloneWorks, and NiCAD achieve 99\% precision when following a similar evaluation methodology. It should also be noted that these three approaches are tuned towards very similar clone detection and show poorer recall for MT3 and WT3/T4 clones, also potentially impacting the precision comparison. Precision results for SourcererCC are mixed: 98\% precision is reported in the original paper by Sajnani et al.\cite{Sajnani2016} but it is 83\% as reproduced by Li et al. \cite{Li2020}. This latter figure, is comparable to the precision of SSCD (for agreed candidates) and worse with respect to our `optimistic' precision score. The precision for Oreo is slightly better than SSCD (90\%) \cite{Saini2018} and precision of CCAligner is worse at 80\%.

Contextualizing our results, scalable recall was our industrial partner's primary goal. Beyond this, they wanted sufficient precision, so that their developers were not spending inordinate amounts of time inspecting false positives. That is, they want to (ideally) find the maximum amount of clones possible and are willing to accept sufficient precision in that context. A hit rate of 1/5 (20\%) was deemed sufficient in this context. The approach substantially exceeds this specified threshold.

\subsubsection{Evaluating SSCD efficiency with BCB}
\label{subsubsec:efficiency_eval}

To compare the time efficiency of SSCD, total execution time is measured; a common practice in CDT evaluation \cite{Saini2018, Sajnani2016}. The total execution time here includes parsing time, source code transformation time, index build time, and search time, which are also reported. Again, this analysis combines actual data (measured directly during our experiments) and estimated data (approximated based on previous research, as reported in the literature). For the latter, the following procedure was used:
\begin{itemize}
    \item First, actual execution time was measured for SourcererCC on the L1 machine, because this technique is routinely reported for comparison.
    \item Li et al. \cite{Li2020} compiled execution times for all the other scalable CDTs. Using these directly is inappropriate due to hardware differences. Instead, it is possible to calculate their difference coefficients with respect to SourcererCC: Oreo is 6.2 times slower for the full BCB, SAGA is 25.7 times faster, CCFinder is 1.4 times faster, and CloneWorks (aggressive) is 4.5 times faster.
    \item Then SourcererCC's actual data is used as a reference point to estimate execution time for CCFinder, CloneWorks (an aggressive configuration was selected, because it has a better execution time), Oreo, and SAGA using the coefficients above. 
    \item For fair comparison with SAGA, we estimate SSCD total execution time if a single GPU was used for inference. Parsing time here is not affected: 51 minute. On 5 GPUs, inference takes 1 hour and 33 minutes: therefore, we estimate the inference time of 7 hours 45 minutes if a single GPU is used. Likewise, index build (3 minutes) and search time (26 minutes) would increase to 2 hours and 25 minutes, combined.
\end{itemize}

The total approximated execution times are shown in Table~\ref{tbl:time_bcb_full}. In column 1 of the table, "(A)" stands for aggressive CloneWorks and in the ``Data source'' column, it means that actual data was used whereas ``E'' means that estimated data was used. As the table shows, 1-GPU SSCD outperforms all CDTs except SAGA: Single GPU-SAGA is estimated to be 4 times faster. While SSCD is more effective and the performance results are comparable if SSCD is executed on 5 GPUs, SAGA too can be scaled to multiple GPUs. Interestingly, for SSCD, inference time adds the most to total execution time on 5 GPUs (54\% of the total), followed by the parsing time (29\% of the total). Index building and search time, combined, make for mere 17\% of the total time.

\begin{table}[!th]
\caption{Total execution time comparison for scalable CDTs}
\label{tbl:time_bcb_full}
\centering
\resizebox{\columnwidth}{!}{
\begin{tabular}{lllr}
\hline
CDT                                       & Machine & Data source & Total time  \\ \hline
\multicolumn{1}{l|}{SSCD (exact, 1 GPU)} &       & E           & 11h 01m \\
\multicolumn{1}{l|}{SSCD (exact, 5 GPUs)} & L1      & A           & 2h 53m      \\
\multicolumn{1}{l|}{SourcererCC (CPU)}          & L1      & A           & 2d 20h 25m  \\
\multicolumn{1}{l|}{CCFinder (CPU)}             &         & E           & 2d 00h 52m  \\
\multicolumn{1}{l|}{CloneWorks (A) (CPU)}       &         & E           & 15h 12m     \\
\multicolumn{1}{l|}{Oreo (CPU)}                 &         & E           & 17d 16h 11m \\
\multicolumn{1}{l|}{SAGA (1 GPU)}                 &         & E           & 2h 40m      \\ \hline
\end{tabular}
}
\end{table}

\subsubsection{Discussion of SSCD comparison to other CDTs}
Answering RQ2, SSCD seems to be more effective than other CDTs used in this evaluation. For the company's datasets its F-score was at least 3 times better than SourcererCC, although the real gain was in precision. On the BCB full dataset, SSCD showed better recall than SAGA for 4/6 clone types and was better for 6/6 clone types than SourcererCC. Precision here was sufficient for the company: comparable to SourcererCC, better than CCAligner, yet worse than SAGA and Oreo.

In terms of execution time, SSCD (in its optimal configuration) is comparable to SAGA, but slower (if of the same order) when a similar hardware configuration is used. SSCD outperforms all other existing CDTs on the three datasets.

\subsubsection{A note on BCB as a Benchmark}
In this study several important insights regarding BCB as a benchmark arose. Most interestingly it seems that BCB has very large clone classes: Recall increased substantially when we changed TopN from 10 to 100 and improved further when we changed it to 250 (not reported). It is unlikely that this is representative of real-world systems: for example, in the company's datasets clone classes are small and a topN of up to 10 is sufficient. Additionally, there are many clones in BCB not explicitly identified in the benchmark (based on our manual inspection of our 'false positives') and researchers should also be aware of that characteristic. 

\section{Threats to Validity}
\label{sec:threats}
The empirical evaluation provided in this paper regarding configuration of SSCD is potentially of limited external validity, in that it is based on the characteristics of the specific dataset employed. For example, topN=100 is based on the prevalence of large clone classes in BCB and setting `code-length' to 128 means that only the starting segments of larger functions are compared. Real code with larger functions may benefit from a code length of 512 and a smaller TopN.

Another validity concern is that part of SSCD's configuration was performed on the company's C/C++ dataset, before being applied to Java (and after being trained on Java). A C/C++ training set would probably improve the relative performance of SSCD on C/C++, but might lead to a less-than-optimal model for subsequent Java-oriented studies, lessening the effectiveness of SSCD in the BCB studies presented.

In comparing SSCD at scale to other CDTs we performed a direct empirical comparison with SourcererCC only. For other comparisons we relied on results reported on/derived from the literature and did not compare its effectiveness with any NN approach. The former leaves open the possibility of discrepancies in machine configurations and empirical protocols that might impact our comparisons. For the latter, we could have directly compared our approach with a pairwise-comparison, NN approach on BCB13 for a more complete evaluation. But, given that BCB13 contains 300,000 code fragments for comparison (10+ LOC methods), that would involve 45 billion comparisons. Conservatively estimating each comparisons at one second, the execution time would be over 520,000 days. We do intend to compare SSCD to Oreo directly on BCB13 though, as it limits potential pairwise comparison significantly.   

Finally, when specifying minimum LOCs, we had to use a proxy-parameter of minimum tokens for SourcererCC. We conservatively estimated 5 tokens per line of code.

\section{Conclusions and the Future Work}
\label{sec:conclusions}
This paper introduces, configures and evaluates SSCD, an approach targeted at effective, scalable detection of clones. It differs from other NN-based approaches in that it does not classify code-pairs into clone/non-clone categories, but instead looks to a segments' nearest neighbours for potential clones.

Overall, SSCD outperforms traditional CDTs in terms of its effectiveness on both the company's and BCB datasets. This is coupled with sufficient precision  and high performance, enabling analysis of sizable codebases (320 MLOC) in a reasonable time for industrial-level contexts. Future work will include a full empirical comparison over other NN/non NN-based CDTs; employing directly-obtained data. 

\section*{Acknowledgment}

This work was supported, in part, by Science Foundation Ireland grant 16/RC/3918.





\bibliographystyle{IEEEtran}
\bibliography{2022-icsme}
\end{document}